\documentclass[preprint,showpacs,preprintnumbers,amsmath,amssymb]{revtex4}

\usepackage{graphicx,color}

\begin{document}

\title{
Finite-distance corrections to the gravitational bending angle of light 
in the strong deflection limit}
\author{Asahi Ishihara}
\author{Yusuke Suzuki}
\author{Toshiaki Ono}
\author{Hideki Asada} 
\affiliation{
Faculty of Science and Technology, Hirosaki University,
Aomori 036-8561, Japan} 
\date{\today}

\begin{abstract} 
Continuing work initiated in an earlier publication 
[Ishihara, Suzuki, Ono, Kitamura, Asada, 
Phys. Rev. D {\bf 94}, 084015 (2016) ], 
we discuss a method of calculating 
the bending angle of light in a static, spherically symmetric 
and asymptotically flat spacetime, especially 
by taking account of the finite distance from a lens object 
to a light source and a receiver. 
For this purpose, we use the Gauss-Bonnet theorem 
to define the bending angle of light, 
such that the definition can be valid also in the strong deflection limit. 
Finally, this method is applied to Schwarzschild spacetime 
in order to discuss also possible observational implications. 
The proposed corrections for Sgr A$^{\ast}$ for instance 
are able to amount to $\sim 10^{-5}$ arcseconds 
for some parameter range, 
which may be within the capability of near-future astronomy, 
while also the correction for the Sun in the weak field limit 
is $\sim 10^{-5}$ arcseconds. 
\end{abstract}


\maketitle

\section{Introduction}
Since 1919 \cite{Eddington}, 
experimental confirmations of the theory of general relativity \cite{GR} 
by measuring the gravitational bending of light by the Sun have been done.  
The gravitational bending of light by other astronomical objects 
has been observed at many times. 
The gravitational lensing 
has been well established as one of the powerful tools in astronomy and cosmology.

The gravitational bending of light plays an important role 
also in the theoretical study of gravity, 
for example on a null structure of a spacetime. 
Hagihara found the analytical solution for light trajectories 
in the gravitational field of Schwarzschild \cite{Hagihara}, 
where the expressions involve incomplete elliptic integrals of the first kind. 
See \cite{Ch} for exact solutions for the light trajectory 
in Schwarzschild and some other black hole solutions. 
See e.g. \cite{MTW,Brumberg} for a review on the light deflection 
in the weak-field approximation of Schwarzschild spacetime. 
A generalized lens equation for the light deflection in Schwarzschild metric 
in the weak-field approximation
and valid for finite distances of source and observer from the lens was
discussed by Zschocke \cite{Zschocke}.

In order to understand a strong gravitational field,  
a more rigorous formulation of the bending angle has been investigated 
\cite{Frittelli, VE2000, Virbhadra, VNC, VE2002, VK2008, ERT, Perlick}. 
For instance, 
strong-field gravitational lensing in a Schwarzschild black hole 
was investigated by Frittelli, Kling and Newman \cite{Frittelli},  
by Virbhadra and Ellis \cite{VE2000} 
and more comprehensively by Virbhadra \cite{Virbhadra}; 
distinctive lensing features of naked singularities 
were studied 
by many authors 
\cite{VNC, VE2002, VK2008, DA, ERT, Perlick, Abe, Toki, Nakajima, Gibbons}. 
Kitamura, Nakajima and Asada proposed a lens model 
in the inverse powers of the distance as $1/r^n$ \cite{Kitamura}, 
such that the Schwarzschild lens, the Ellis wormhole lens 
and a gravitational lens associated with exotic matter (or energy) 
that might follow a non-standard equation of state 
can be discussed in a unified manner of describing 
these models as a one parameter family 
\cite{Tsukamoto,Izumi,Kitamura2014,Nakajima2014}. 
See also Tsukamoto et al. (2015) \cite{Tsukamoto2014} 
for a possible connection between this inverse power model and 
the Tangherlini solution to the higher-dimensional Einstein equation. 

Gibbons and Werner (2008) proposed an alternative way of deriving 
the deflection angle of light \cite{GW2008}. 
They used the Gauss-Bonnet theorem to a spatial domain 
described by the optical metric, 
where they assumed that the source and receiver are located 
at an asymptotic region. 
By extending Gibbons and Werner's idea, 
Ishihara et al. have recently investigated finite-distance corrections 
\cite{Ishihara}. 
However, such works \cite{GW2008,Ishihara} are limited 
within the small deflection limit. 

Since the pioneering work by Darwin \cite{Darwin}, on the other hand, 
the strong deflection limit has attracted a lot of interests 
(See e.g. \cite{Tsukamoto,Bozza,Iyer,Bozza+}), 
mainly because we expect that 
the recent progress in astronomical instruments 
will soon enable us to detect such strong deflection phenomena. 

Therefore, the main purpose of the present paper is 
to extend the earlier work \cite{Ishihara}, 
especially in order to examine finite-distance corrections 
in the strong deflection limit. 
We shall discuss their implications for possible observations. 

Throughout this paper, we use the unit of $G=c=1$. 
In the following, the observer may be called the receiver 
in order to avoid a confusion between $r_O$ and $r_0$ by using $r_R$.

\section{Gravitational deflection of light and Gauss-Bonnet theorem}

\subsection{Notation}
Following Ishihara et al. \cite{Ishihara}, 
the present paper considers a static and spherically symmetric (SSS) spacetime. 
The SSS spacetime can be described as (cf. Eq.(23.3) in \cite{MTW}) 
\begin{eqnarray}
ds^2 &=& g_{\mu\nu} dx^{\mu} dx^{\nu} 
\nonumber\\
&=& -A(r) dt^2 + B(r) dr^2 + C(r) d\Omega^2 , 
\label{ds2-SSS-AB}
\end{eqnarray}
where 
$d\Omega^2 \equiv d\theta^2 + \sin^2\theta d\phi^2$. 
This metric form allows us to consider 
a wormhole solution with a throat as well as a black hole spacetime. 
If we choose $C(r) = r^2$, then, $r$ denotes the circumference radius.  
Without the loss of generality, henceforth, we choose 
the photon orbital plane as the equatorial plane 
($\theta = \pi/2$). 
As usual, we define the impact parameter of the light ray as 
\begin{eqnarray}
b &\equiv& \frac{L}{E} 
\nonumber\\
&=& \frac{C(r)}{A(r)} \frac{d\phi}{dt} , 
\label{b}
\end{eqnarray}
such that we can obtain the orbit equation as 
\begin{eqnarray}
\left( \frac{dr}{d\phi} \right)^2 
+ \frac{C(r)}{B(r)} 
= \frac{[C(r)]^2}{b^2 A(r)B(r)} . 
\label{orbiteq}
\end{eqnarray}

Light rays satisfy the null condition as $ds^2 = 0$, 
which is rearranged as, via Eq. (\ref{ds2-SSS-AB}), 
\begin{eqnarray}
dt^2 &=& \gamma_{IJ} dx^I dx^J 
\nonumber\\
&=& \frac{B(r)}{A(r)} dr^2 + \frac{C(r)}{A(r)} d\phi^2 ,  
\label{gamma}
\end{eqnarray}
where $I$ and $J$ denote $1$ and $2$. 
$\gamma_{IJ}$ is called the optical metric. 
This optical metric defines a two-dimensional Riemannian space 
(denoted as $M^{\mbox{opt}}$), 
in which the light ray is a spatial geodetic curve.

In $M^{\mbox{opt}}$, 
let $\Psi$ denote the angle of the light ray measured from 
the radial direction. 
We obtain  
\begin{eqnarray}
\cos \Psi = 
\frac{b \sqrt{A(r)B(r)}}{C(r)} \frac{dr}{d\phi} , 
\label{cosPsi2}
\end{eqnarray}
which is rewritten in a more convenient form as 
\begin{equation}
\sin\Psi = \frac{b \sqrt{A(r)}}{\sqrt{C(r)}} , 
\label{sinPsi}
\end{equation}
where we used Eq. (\ref{orbiteq}).

Let $\Psi_R$ and $\Psi_S$ denote the angles that are measured 
at the receiver position (R) and the source position (S), respectively. 
Let $\phi_{RS} \equiv \phi_R - \phi_S$ denote 
the coordinate separation angle
between the receiver and source. From the three angles 
$\Psi_R$, $\Psi_S$ and $\phi_{RS}$, 
let us define \cite{Ishihara} 
\begin{equation}
\alpha \equiv \Psi_R - \Psi_S + \phi_{RS} . 
\label{alpha}
\end{equation}

Eq.(\ref{alpha}) has a geometrically invariant meaning 
as follows \cite{Ishihara} . 

\begin{figure}
\includegraphics[width=12cm]{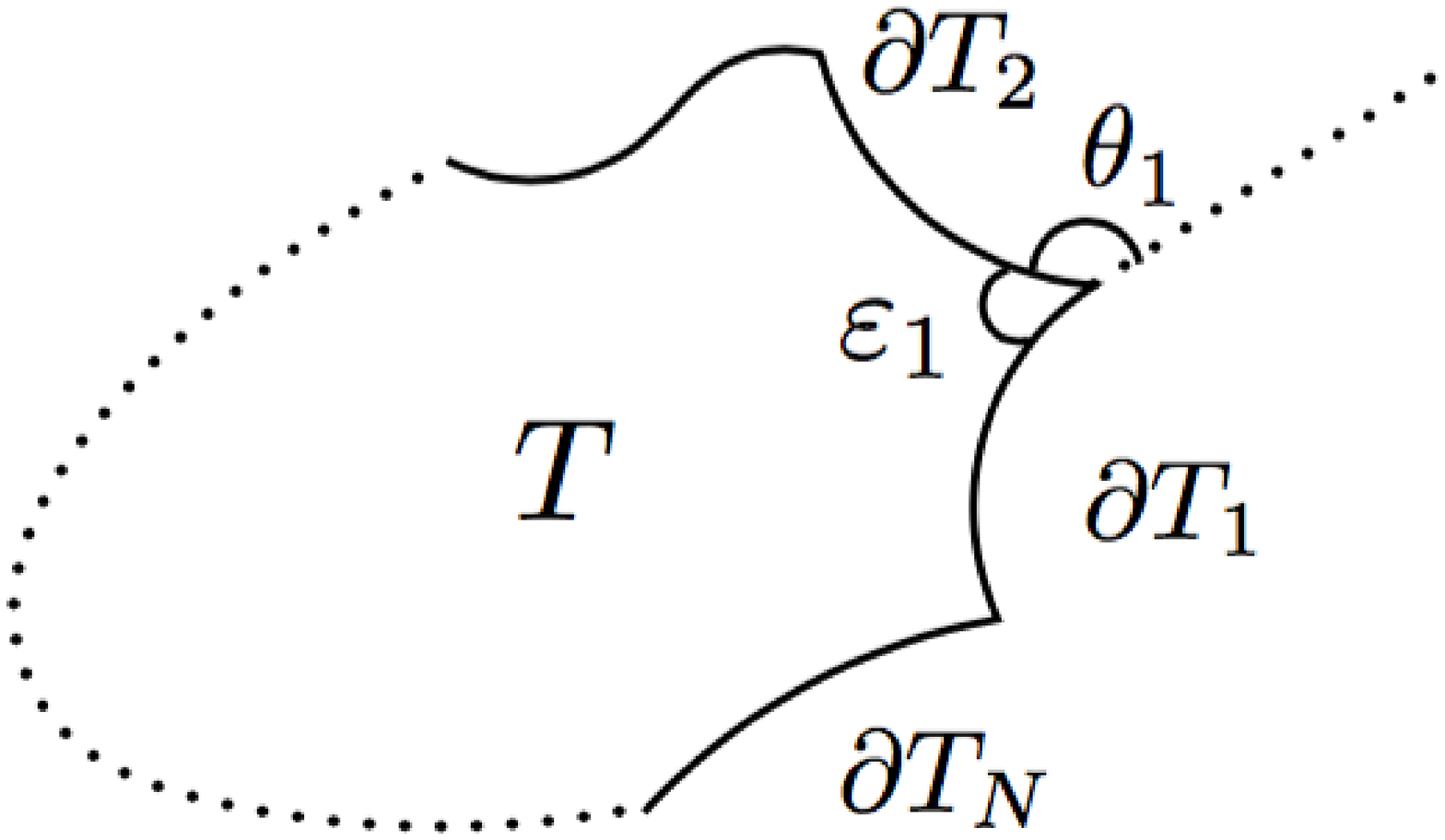}
\caption{
Schematic figure for the Gauss-Bonnet theorem. 
}
\label{fig-GB}
\end{figure}

Suppose that 
$T$ is a two-dimensional orientable surface with boundaries $\partial T_a$ 
($a=1, 2, \cdots, N$) that 
are differentiable curves. 
See Figure \ref{fig-GB}. 
Let the jump angles between the curves be $\theta_a$ 
($a=1, 2, \cdots, N$). 
Then, the Gauss-Bonnet theorem can be expressed as 
\cite{GB-theorem}
\begin{eqnarray}
\iint_{T} K dS + \sum_{a=1}^N \int_{\partial T_a} \kappa_g d\ell + 
\sum_{a=1}^N \theta_a = 2\pi , 
\label{localGB}
\end{eqnarray}
where 
$K$ denotes the Gaussian curvature of 
the surface $T$, 
$dS$ is the area element of the surface, 
$\kappa_g$ means the geodesic curvature of $\partial T_a$, 
and $\ell$ is the line element along the boundary. 
The sign of the line element is chosen such that it is 
compatible with the orientation of the surface. 

\begin{figure}
\includegraphics[width=10cm]{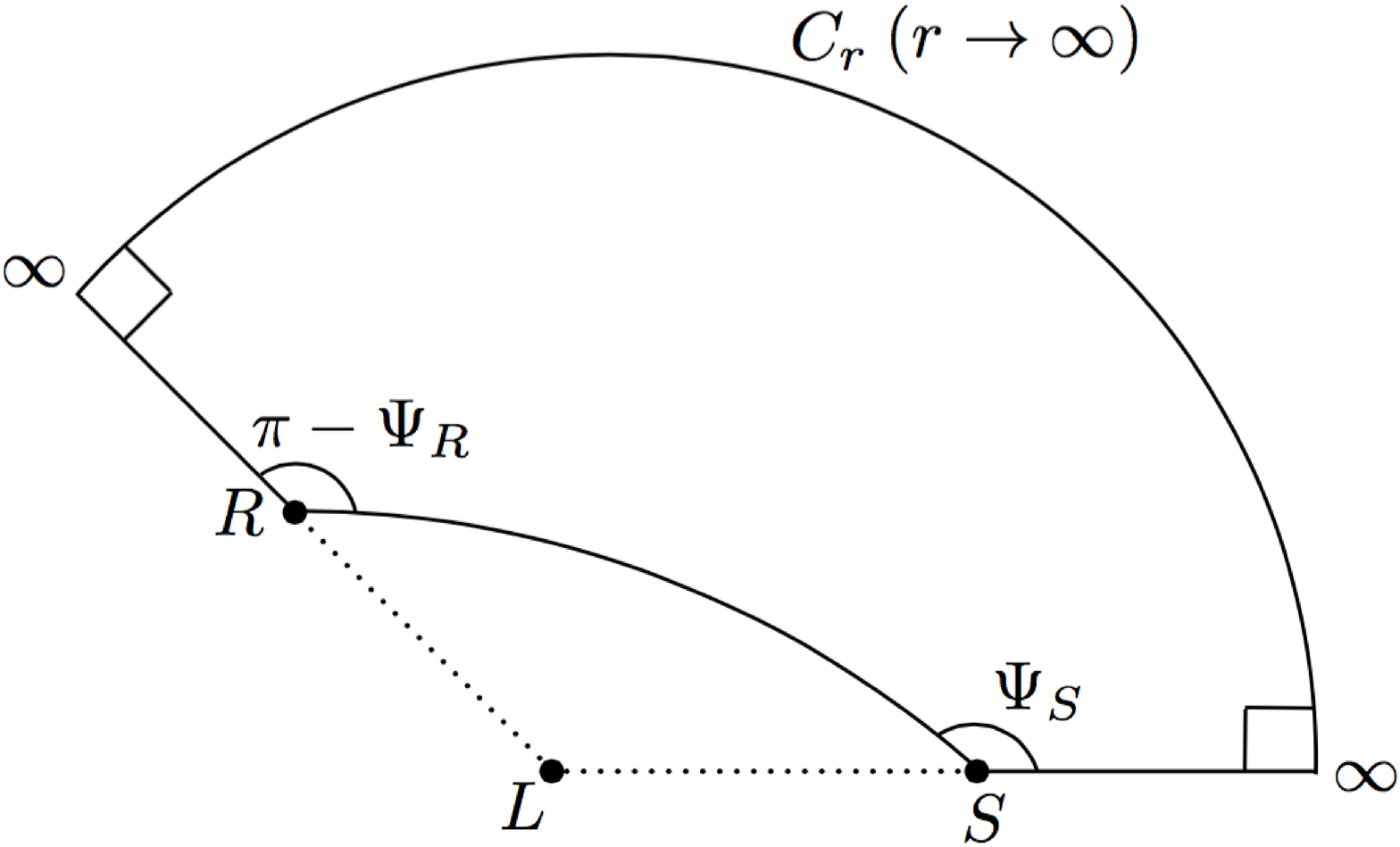}
\caption{
Quadrilateral ${}^{\infty}_{R}\Box^{\infty}_{S}$
embedded in a curved space in $M^{\mbox{opt}}$. 
}
\label{fig-Box}
\end{figure}

Let us consider a quadrilateral 
${}^{\infty}_{R}\Box^{\infty}_{S}$, 
which consists of the spatial curve for the light ray, 
two outgoing radial lines from R and from S 
and a circular arc segment $C_r$ 
of coordinate radius $r_C$ ($r_C \to \infty$) 
centered at the lens 
which intersects the radial lines  
through the receiver or the source. 
See Figure \ref{fig-Box}. 
Henceforth, we restrict ourselves within the asymptotically flat spacetime, 
for which 
$\kappa_g \to 1/r_C$ and $d\ell \to r_C d\phi$ 
as $r_C \to \infty$ (See e.g. \cite{GW2008}). 
Hence, 
$\int_{C_r} \kappa_g d\ell \to \phi_{RS}$.  
Applying this result to the Gauss-Bonnet theorem 
for ${}^{\infty}_{R}\Box^{\infty}_{S}$, 
we obtain 
\begin{eqnarray}
\alpha 
&=& \Psi_R - \Psi_S + \phi_{RS} 
\nonumber\\
&=& - \iint_{{}^{\infty}_{R}\Box^{\infty}_{S}} K dS . 
\label{alpha-K}
\end{eqnarray}
Eq. (\ref{alpha-K}) shows that $\alpha$ 
is invariant in differential geometry 
and $\alpha$ is well-defined even if 
$L$ is a singularity point. 
Moreover, it follows that $\alpha=0$ in Euclidean space. 

Eq. (\ref{alpha-K}) recovers 
the deflection angle of light in the far limit of 
the source and the receiver as 
\begin{equation}
\alpha_{\infty} = 2 \int_0^{u_0} \frac{du}{\sqrt{F(u)}} - \pi , 
\label{alpha-infty}
\end{equation}
where 
$u$ is the inverse of $r$, 
$u_0$ is the inverse of the closest approach 
(often denoted as $r_0$) and 
$F(u)$ is defined as 
\begin{eqnarray}
F(u) \equiv \left( \frac{du}{d\phi} \right)^2 , 
\label{F}
\end{eqnarray}
which can be computed from Eq. (\ref{orbiteq}). 

Next, we consider another case that 
the distance from the source to the receiver is finite 
because every observed stars and galaxies are 
located at finite distance from us 
(e.g., at finite redshift in cosmology) 
and the distance is much larger than the size of the lens. 
Let $u_R$ and $u_S$ denote the inverse of 
$r_R$ and $r_S$, respectively, 
where $r_R$ and $r_S$ are finite. 
Eq. (\ref{alpha}) becomes \cite{Ishihara}
\begin{equation}
\alpha = \int_{u_R}^{u_0} \frac{du}{\sqrt{F(u)}} 
+ \int_{u_S}^{u_0} \frac{du}{\sqrt{F(u)}} 
+\Psi_R - \Psi_S . 
\label{alpha-finite}
\end{equation}

\section{Extension to a strong deflection limit}
In the previous section for the weak deflection, 
a spatial curve from the source to the receiver is simple. 
In the strong deflection limit, however, a spatial curve 
from the source to the receiver may have a winding number 
that can exceed unity. 
Therefore, the spatial curve may have intersection points. 
We split the curve into pieces, such that each of the pieces 
can have no intersection. 

\subsection{Loops of the curve for the photon orbit}
\begin{figure}
\includegraphics[width=14cm]{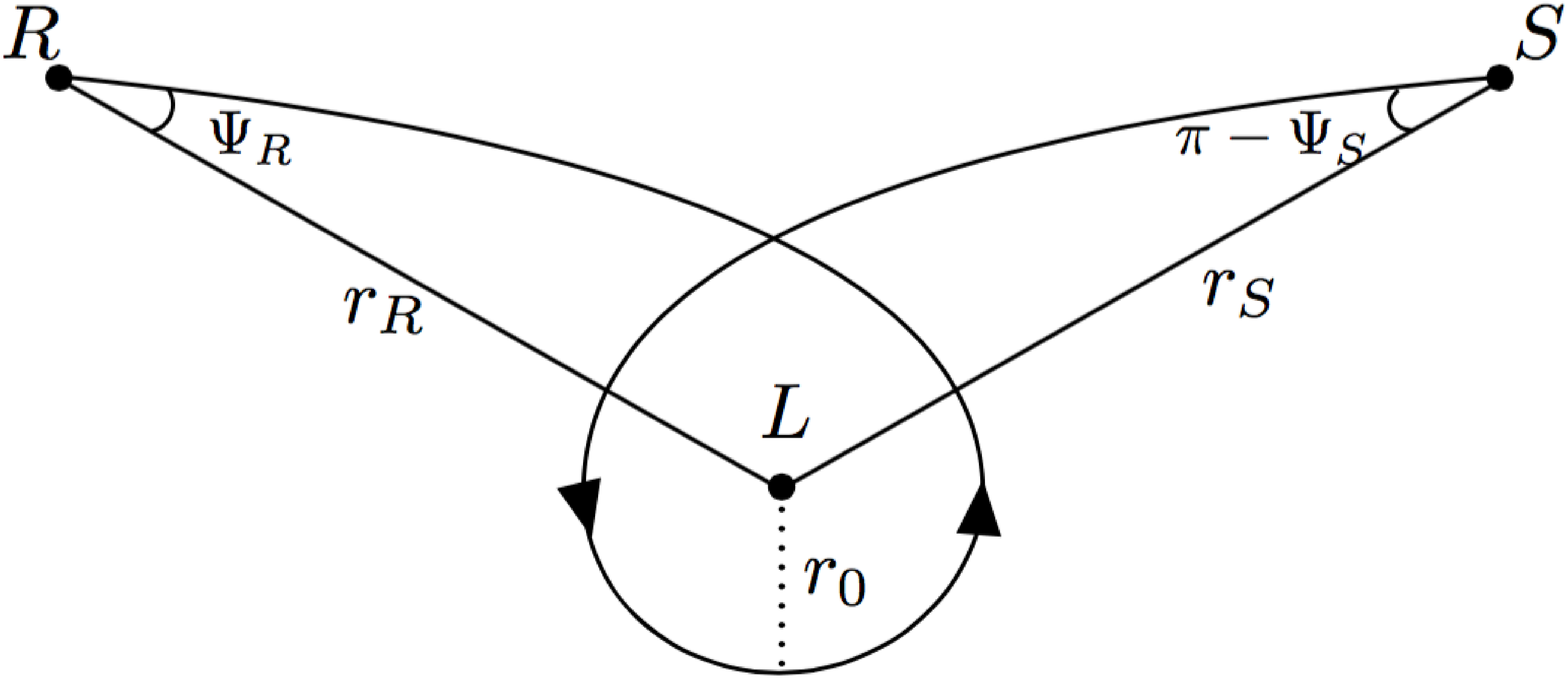}
\caption{
One-loop diagram for the photon trajectory 
in $M^{\mbox{opt}}$. 
}
\label{fig-oneloop}
\end{figure}

\begin{figure}
\includegraphics[width=10cm]{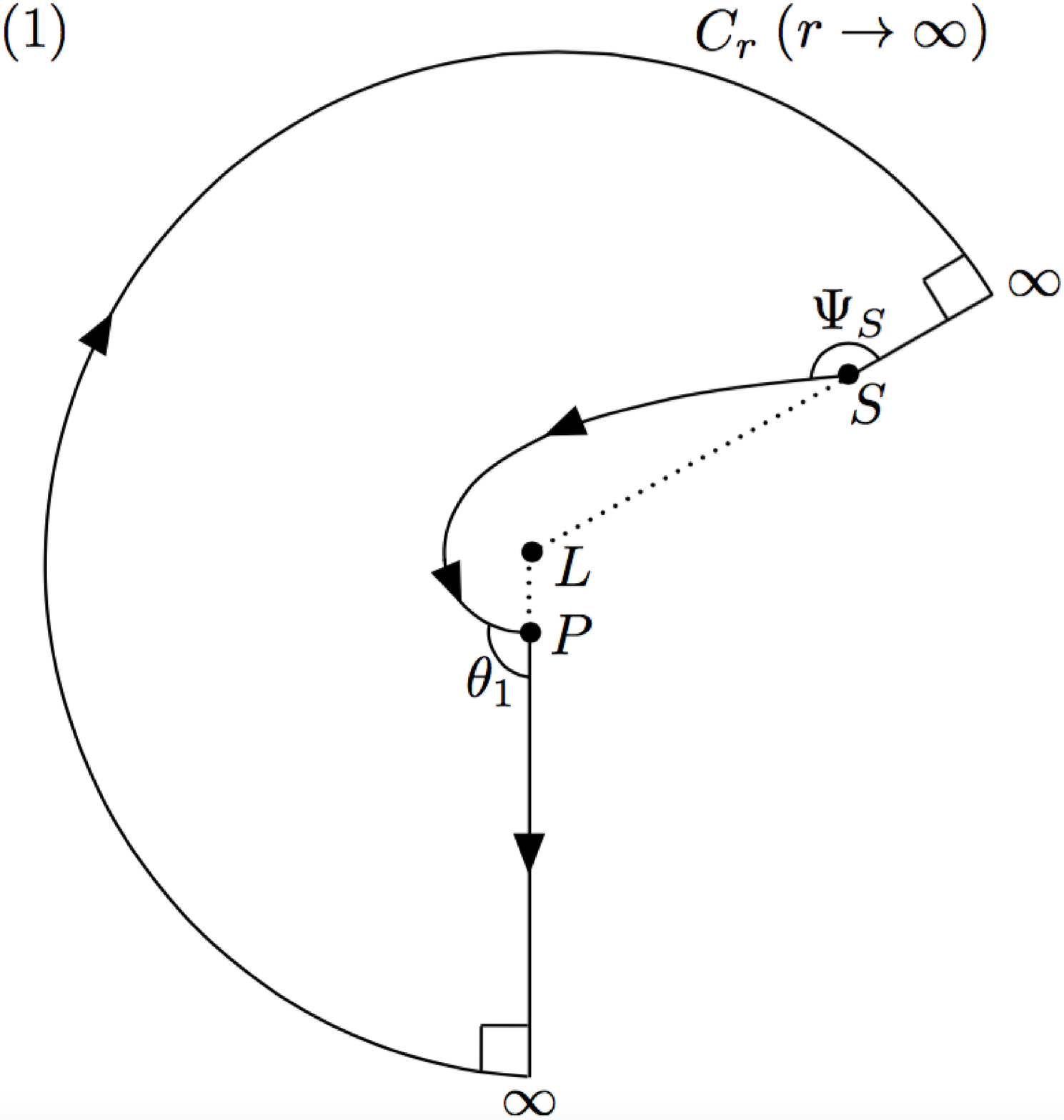}
\includegraphics[width=10cm]{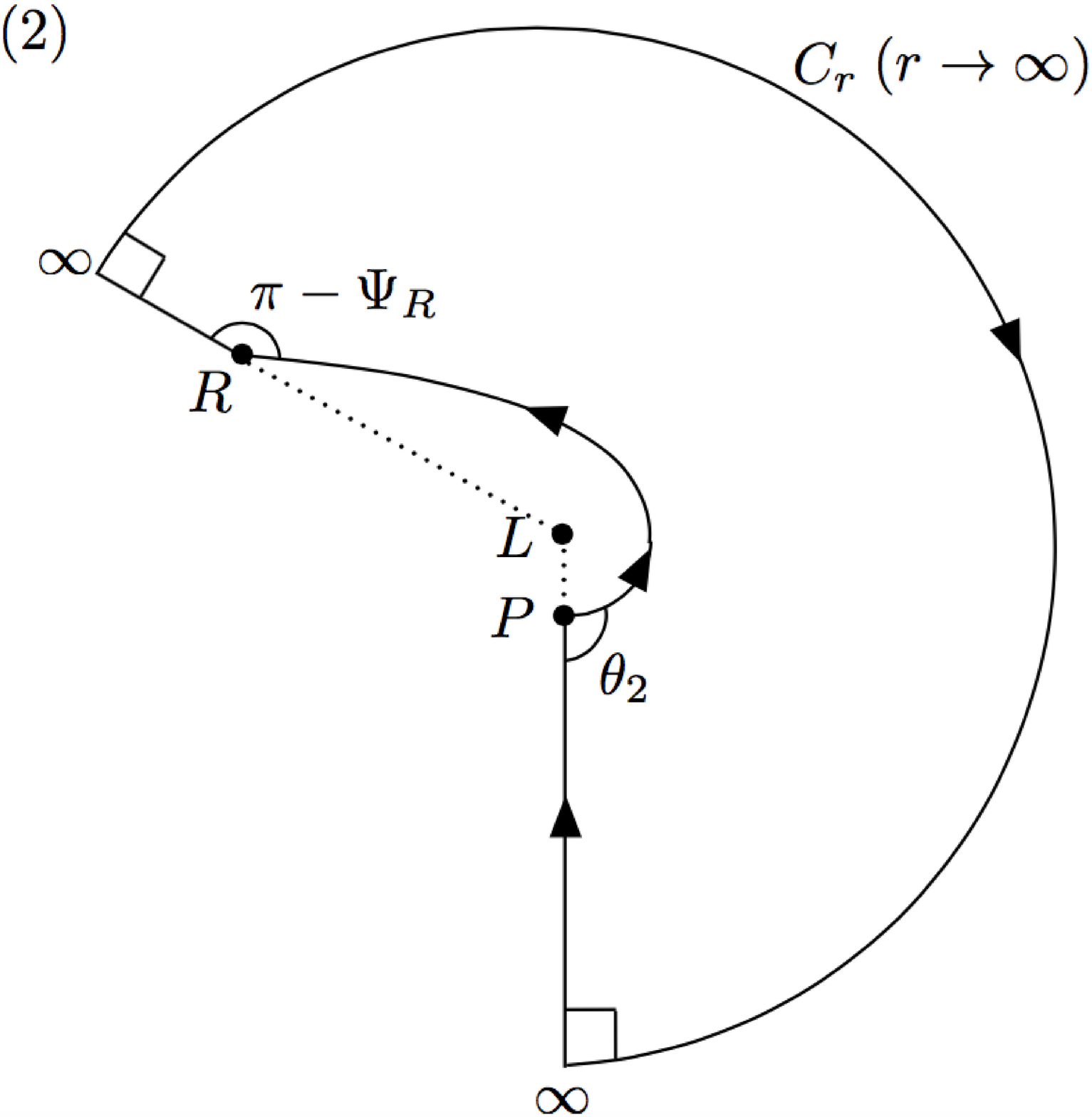}
\caption{ 
Two quadrilaterals from the photon orbit in Figure \ref{fig-oneloop}. 
They are embedded in a non-Euclidean space. 
}
\label{fig-oneloop2}
\end{figure}

We begin with the case of one loop for its simplicity. 
Please see Figure \ref{fig-oneloop}. 
First, we examine the two quadrilaterals (1) and (2) 
in Figure \ref{fig-oneloop2}, 
which can be constructed by imagining an auxiliary point (P) 
that may be the periastron and 
then by assuming auxiliary outgoing radial lines 
(solid line in the figure) from 
the point P in the quadrilaterals (1) and (2). 
In fact, it does not matter that the point P is chosen 
as the periastron. 
Note that the direction of the two auxiliary lines in (1) and (2) 
is opposite and thus the two auxiliary lines cancel out 
to make no contributions to $\alpha$ 
in the sum of the quadrilaterals (1) and (2).  
Let $\theta_1$ and $\theta_2$ denote the inner angle 
at the point P in the quadrilateral (1) 
and that in the quadrilateral (2), respectively. 
We find $\theta_1 + \theta_2 = \pi$, 
because the line from the source to the receiver is a geodesic curve 
and the point P lies on the geodesic. 

For each orientable quadrilateral in Figure \ref{fig-oneloop2}, 
the method in the previous section can be applied as it is. 
In the similar manner to deriving Eq. (\ref{alpha-K}),  
we obtain 
\begin{align}
\alpha^{(1)} &= 
(\pi - \theta_1) - \Psi_S + \phi_{RS}^{(1)} , 
\nonumber\\
\alpha^{(2)} &= 
\Psi_R - \theta_2 + \phi_{RS}^{(2)} ,   
\end{align}
where 
the coordinate angle difference $\phi_{RS}$ 
is split into two parts, 
$\phi_{RS}^{(1)}$ for the quadrilateral (1) and 
$\phi_{RS}^{(2)}$ for the other quadrilateral (2). 

If $r_S = r_R$, then, there is a reflection symmetry 
between the quadrilaterals (1) and (2) and 
$\phi_{RS}^{(1)} = \phi_{RS}^{(2)} = \phi_{RS}/2$. 
Otherwise, $\phi_{RS}^{(1)}$ does not always equal to $\phi_{RS}^{(2)}$. 
In any case, however, $\phi_{RS}^{(1)} + \phi_{RS}^{(2)} = \phi_{RS}$. 
$\Psi_S$ and $(\pi - \Psi_R)$ are 
the inner angles at $S$ and $R$, respectively, 
where we should remember that $\Psi_R$ 
is an angle measured from the outgoing radial line. 
We thus obtain 
\begin{align}
\alpha &= \alpha^{(1)} + \alpha^{(2)} 
\nonumber\\
&= \Psi_R - \Psi_S + \phi_{RS} , 
\end{align}
where we use 
$\theta_1 + \theta_2 = \pi$ 
and
$\phi_{RS}^{(1)} + \phi_{RS}^{(2)} = \phi_{RS}$. 
This result takes the same form as Eq. (\ref{alpha}), 
but we should note that it is derived for one loop case. 

\begin{figure}
\includegraphics[width=14cm]{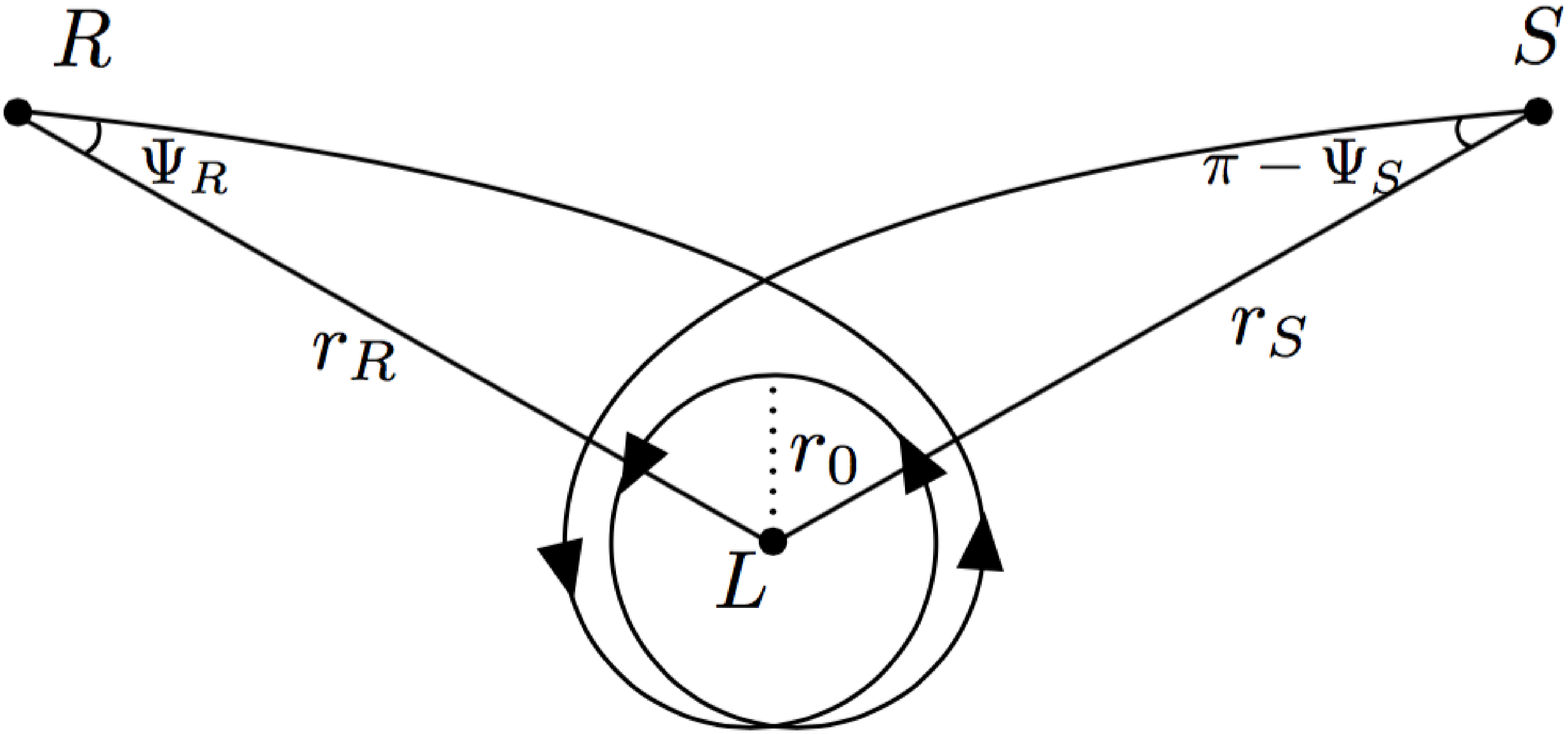}
\caption{ 
Two-loop diagram for the photon trajectory 
in $M^{\mbox{opt}}$. 
}
\label{fig-twoloop}
\end{figure}

\begin{figure}
\includegraphics[width=7cm]{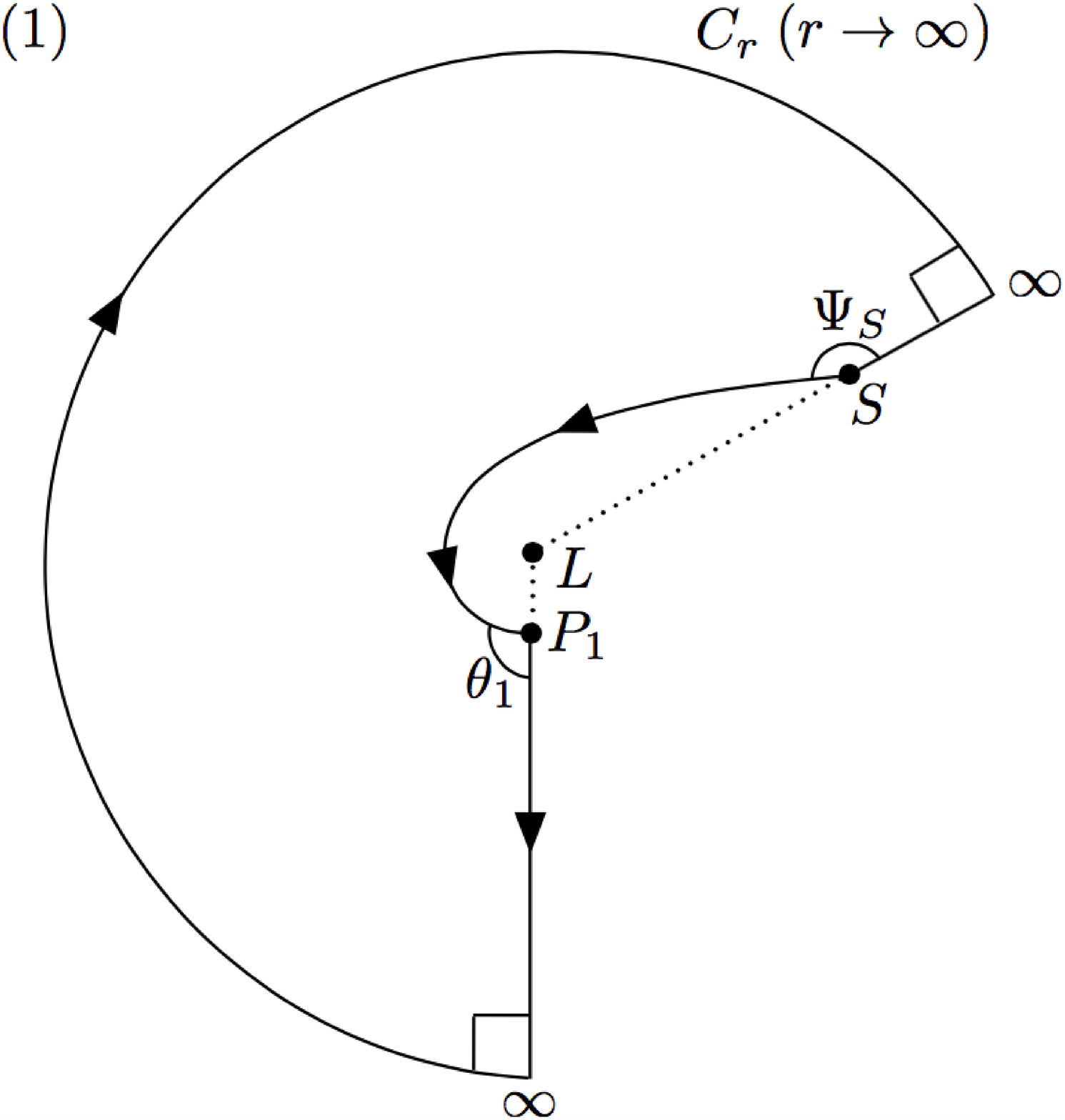}
\includegraphics[width=8cm]{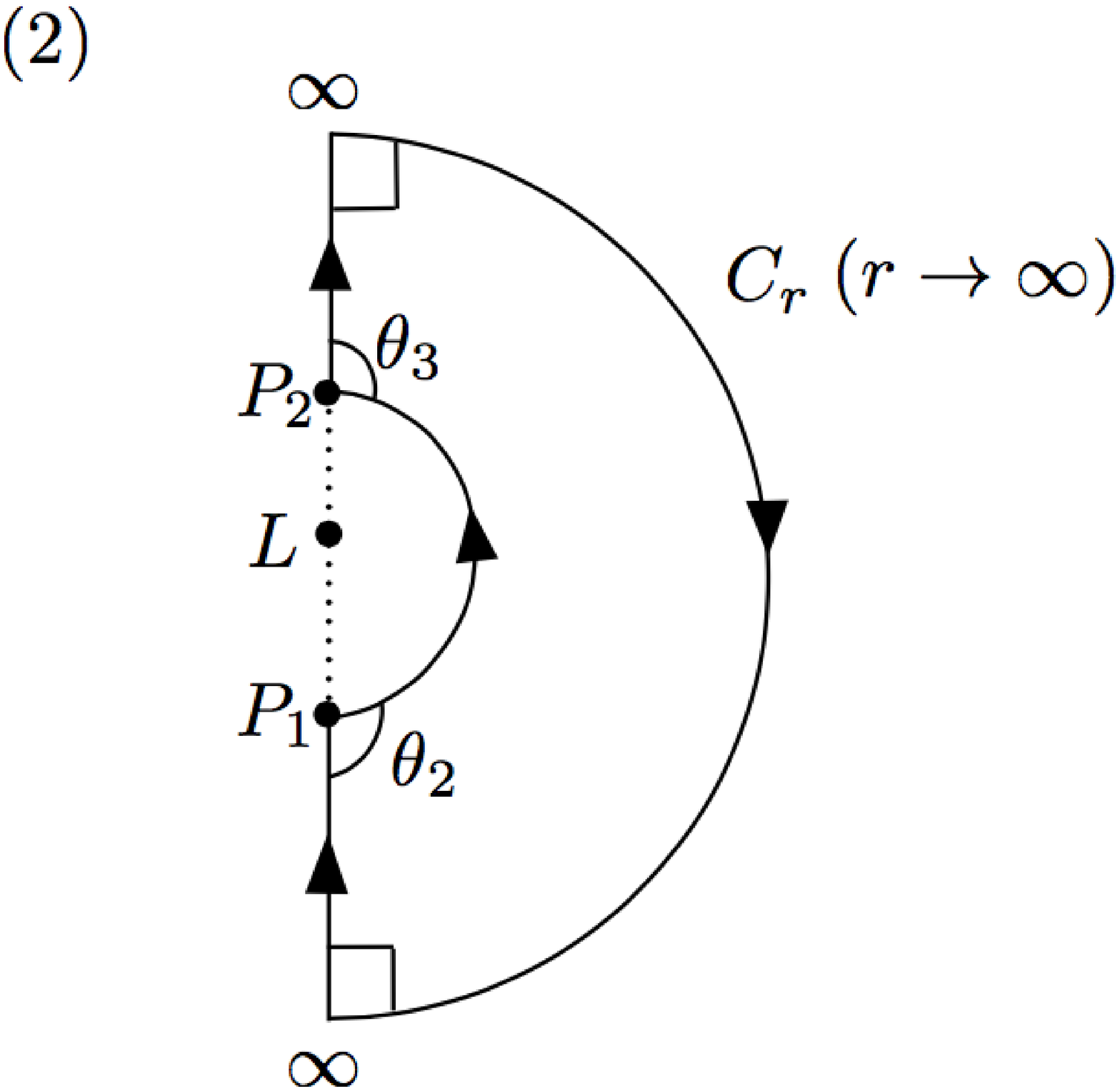}
\includegraphics[width=6cm]{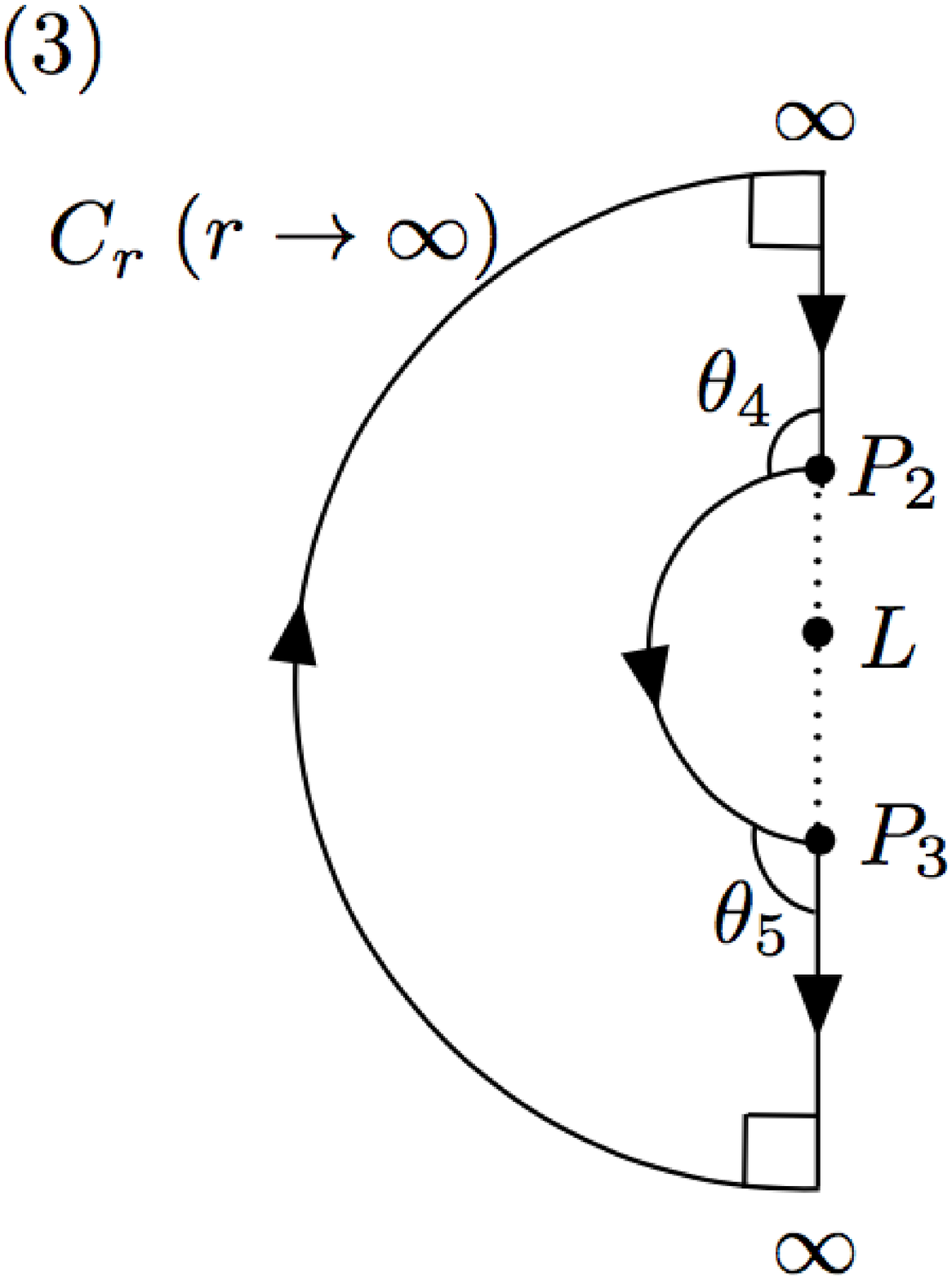}
\includegraphics[width=7cm]{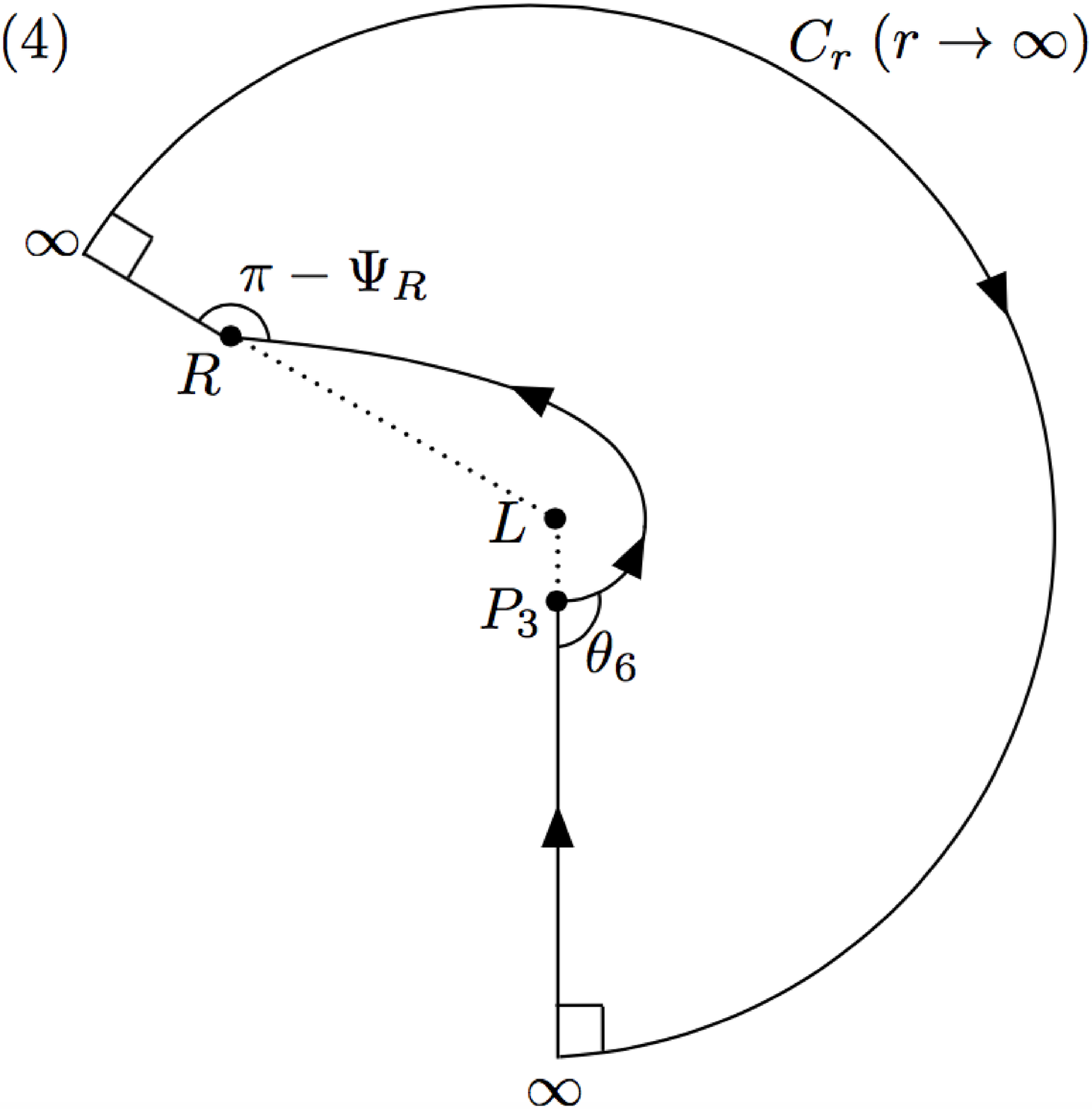}
\caption{ 
Four quadrilaterals (1)-(4) 
in a non-Euclidean space $M^{\mbox{opt}}$. 
They are constructed from the two-loop diagram 
for the photon orbit in Figure \ref{fig-twoloop}. 
}
\label{fig-twoloop2}
\end{figure}

Next, we consider a case of two loops as shown by Figure \ref{fig-twoloop}, 
for which we draw auxiliary lines to 
split the configuration into four quadrilaterals 
(See Figure \ref{fig-twoloop2}). 
For each quadrilateral, we obtain 
\begin{align}
\alpha^{(1)} &= 
(\pi - \theta_1) - \Psi_S + \phi_{RS}^{(1)} , 
\nonumber\\
\alpha^{(2)} &= 
(\pi- \theta_3) - \theta_2 + \phi_{RS}^{(2)} ,
\nonumber\\
\alpha^{(3)} &= 
(\pi- \theta_5) - \theta_4 + \phi_{RS}^{(3)} ,
\nonumber\\
\alpha^{(4)} &= 
\Psi_R - \theta_6 + \phi_{RS}^{{(4)}} ,   
\end{align}
where 
$\phi_{RS}^{(1)} + \phi_{RS}^{(2)} + \phi_{RS}^{(3)} + \phi_{RS}^{(4)} 
= \phi_{RS}$. 
We thus obtain 
\begin{align}
\alpha &= \alpha^{(1)} + \alpha^{(2)} + \alpha^{(3)} + \alpha^{(4)}
\nonumber\\
&= \Psi_R - \Psi_S + \phi_{RS} , 
\label{alpha-2loop}
\end{align}
where we use 
$\theta_1 + \theta_2 = \theta_3 + \theta_4 = \theta_5 + \theta_6 = \pi$.  
While Eq. (\ref{alpha-2loop}) is derived for the two-loop case, 
it reveals again the form of Eq. (\ref{alpha}). 
Note that one loop, from which the quadrilaterals (2) and (3) 
can be constructed, 
makes the contribution to $\alpha$ 
only in terms of $\phi_{RS}^{(2)} + \phi_{RS}^{(3)}$.

\begin{figure}
\includegraphics[width=12cm]{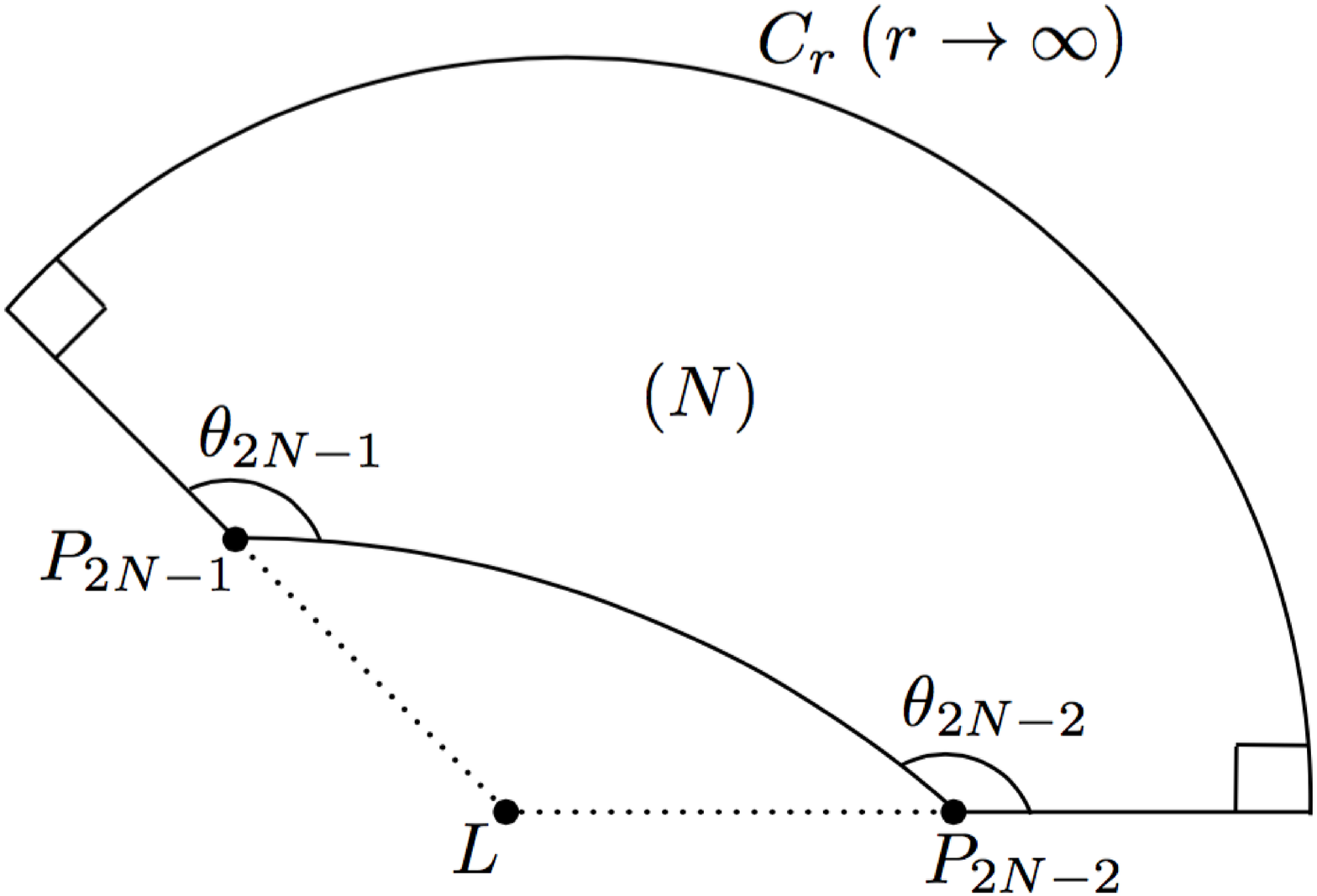}
\caption{ 
One of quadrilaterals from a photon orbit with any winding number. 
This can be used in order to prove by induction that Eq. (\ref{alpha}) 
holds for any loop number case. 
}
\label{fig-anyloop}
\end{figure}

Finally, we consider any winding number $W$. 
We construct $2W$ quadrilaterals, for which 
the inner angles at finite distance from $L$ 
are denoted as $\theta_0, \cdots, \theta_{2W}$ 
in order from $S$ to $R$. 
Here, $\theta_0 = \Psi_S$ and $\theta_{2W} = \pi - \Psi_R$. 
See Figure \ref{fig-anyloop} for one of the quadrilaterals. 
Any pair of neighboring quadrilaterals (N) and (N+1) 
makes the contribution to $\alpha$ only by 
$\phi_{RS}^{(N)} + \phi_{RS}^{(N+1)}$, 
because $\theta_{2N-1} + \theta_{2N} = \theta_{2N+1} + \theta_{2N+2} = \pi$ 
and the auxiliary lines cancel out.  
By induction, therefore, 
Eq. (\ref{alpha}) can be derived for any winding number.

Eq. (\ref{alpha}), which is equivalent to Eq. (\ref{alpha-finite}) 
by using the orbit equation, 
is rearranged as 
\begin{align}
\alpha&=\Psi_{R}-\Psi_{S}+\phi_{RS}
\notag\\
&=\Psi_{R}-\Psi_{S}
+\int^0_{u_R}\frac{du}{\sqrt{F(u)}}
+\int^0_{u_S}\frac{du}{\sqrt{F(u)}}
+2\int^{u_0}_0\frac{du}{\sqrt{F(u)}} . 
\label{alpha-full}
\end{align}

The finite-distance correction to the deflection angle of light, 
denoted as $\delta\alpha$,  
is the difference between the asymptotic deflection angle 
and the deflection angle for the finite distance case. 
It is expressed as  
\begin{align}
\delta\alpha = \alpha - \alpha_{\infty}  . 
\label{delta-alpha}
\end{align}
Substituting Eqs. (\ref{alpha-infty}) and (\ref{alpha-full}) 
into the right-hand side of Eq. (\ref{delta-alpha}) 
and rearranging it, 
we obtain 
\begin{align}
\delta\alpha = 
(\Psi_{R}-\Psi_{S}+\pi) + \int^0_{u_R}\frac{du}{\sqrt{F(u)}}
+\int^0_{u_S}\frac{du}{\sqrt{F(u)}} . 
\label{delta-alpha2}
\end{align}

This expression suggests two origins of the finite-distance corrections. 
One origin is the angles  
$\Psi_{R}$ and $\Psi_{S}$ 
that are defined at the receiver and the source at finite distance, 
where the space is non-Euclidean 
\cite{comment-Bozza+}. 
The other origin is the two path integrals, 
one from the receiver position to the spatial infinity 
and the other from the source to the infinity. 
Therefore, if both the receiver and the source are in the weak field region 
as is common in astronomy, 
the finite-distance correction comes only from the weak field region 
but not from the strong field region, 
even if the light ray passes near the photon sphere 
$(r_0 \sim r_{ph})$, where $r_{ph}$ denotes the photon sphere radius. 
This is quite reasonable.

\subsection{Approximations}
As a concrete example, henceforth, we focus on the Schwarzschild black hole 
with mass $M$. 
Then, Eq. (\ref{F}) becomes 
\begin{equation}
F(u) = \frac{1}{b^2} - u^2 + 2M u^3 . 
\label{F-Sch}
\end{equation} 

Eq. (\ref{alpha-full}) with $F(u)$ given by Eq. (\ref{F-Sch}) 
can be solved analytically but 
leads to cumbersome expressions involving incomplete elliptic 
integrals of the first kind. 
In the case that the source and the receiver are far from the lens 
($r_S \gg b, r_R \gg b$) 
but the light ray passes near the photon sphere ($r_0 \sim 3M$), 
Eq. (\ref{alpha-full}) can be  
approximated and simplified considerably as 
\begin{align}
\alpha=
&\frac{2M}{b}\left[\sqrt{1-b^2u_R^2}+\sqrt{1-b^2u_S^2}-2\right]
\notag\\
&+2\log\left(\frac{12(2-\sqrt{3})r_0}{r_0-3M}\right)-\pi
\notag\\
&+O\left(\frac{M^2}{r_R{}^2},\frac{M^2}{r_S{}^2},1-\frac{3M}{r_0}\right) ,
\label{alpha-approx}
\end{align}
where the logarithmic term \cite{Iyer} was used 
for the last term of Eq. (\ref{alpha-full}). 
Here, the leading terms in $\Psi_R$ and $\Psi_S$ 
cancel out with the terms coming from the integrals. 
Therefore, $\Psi_R$ and $\Psi_S$ do not appear 
in the final expression of Eq. (\ref{alpha-approx}). 
See Appendix for more details. 

As stated above, it is natural that the logarithmic term due to 
the strong field is 
independent of 
finite-distance corrections 
such as a multiplication by $\sqrt{1 - (b u_S)^2}$. 
By chance, 
$\delta\alpha$ for the strong deflection limit (See Eq. (\ref{alpha-full})) 
is the same as that for the weak deflection case 
(See e.g.  Eq. (29) in \cite{Ishihara}). 
This suggests that 
the finite-distance correction for the strong deflection limit 
is of the same order as  
\begin{align}
\delta\alpha \sim 
O\left(\frac{M b}{r_S{}^2} + \frac{M b}{r_R{}^2}\right) , 
\label{order-alpha}
\end{align}
for the weak field case (e.g. \cite{Ishihara}). 
Namely, the correction is linear in the impact parameter. 
This implies that 
the finite-distance correction for the weak deflection case (large $b$)
is larger than that in the strong deflection limit (small $b$), 
if the other parameters are fixed. 
In the next section, we shall discuss in more detail.

\section{Possible observational candidates}
\subsection{Gravitational bending of light by the Sun} 
We assume that an observer at the Earth sees 
the light bending by the solar mass, while 
the source is practically at the asymptotic region. 
If the light ray passes near the solar surface, 
Eq. (\ref{order-alpha}) implies that 
the finite-distance correction to this case is 
of the order of 
\begin{align}
\delta\alpha 
&\sim 
\frac{M b}{r_R{}^2} 
\nonumber\\
&\sim 
10^{-5} \mbox{arcsec.} 
\times 
\left(\frac{M}{M_{\odot}}\right) 
\left(\frac{b}{R_{\odot}}\right) 
\left(\frac{1 \mbox{AU}}{r_R}\right)^2 , 
\label{alpha-Sun}
\end{align}
where $4M_{\odot}/R_{\odot} \sim 1.75 \,\mbox{arcsec.}$, 
and $R_{\odot}$ denotes the solar radius. 
Note that Eq. (\ref{alpha-Sun}) comes from Eq. (29) in 
Ref. \cite{Ishihara} for the weak-field limit 
and therefore it is independent of Eq. (\ref{alpha-approx}) 
for the strong deflection limit. 

This correction of $\sim 10^{-5}$ arcsec. 
($= 10$ micro arcseconds), 
is close to the angular accuracy 
within the capability of near-future astronomy. 
For instance, a current astrometry space mission Gaia \cite{Gaia} 
and a future one JASMINE 
(Japan Astrometry Satellite Mission for Infrared Exploration) 
\cite{JASMINE} 
are expected to 
approach nearly ten micro arcseconds, 
though the solar direction is too bright 
and even dangerous for these telescopes. 
In order to measure the above finite-distance correction, 
a specially dedicated instrument such as a corona graph 
might be needed 
or a total solar eclipse for an astrometry satellite 
could be used along the original idea by Arthur Eddington.

\begin{figure}
\includegraphics[width=10cm]{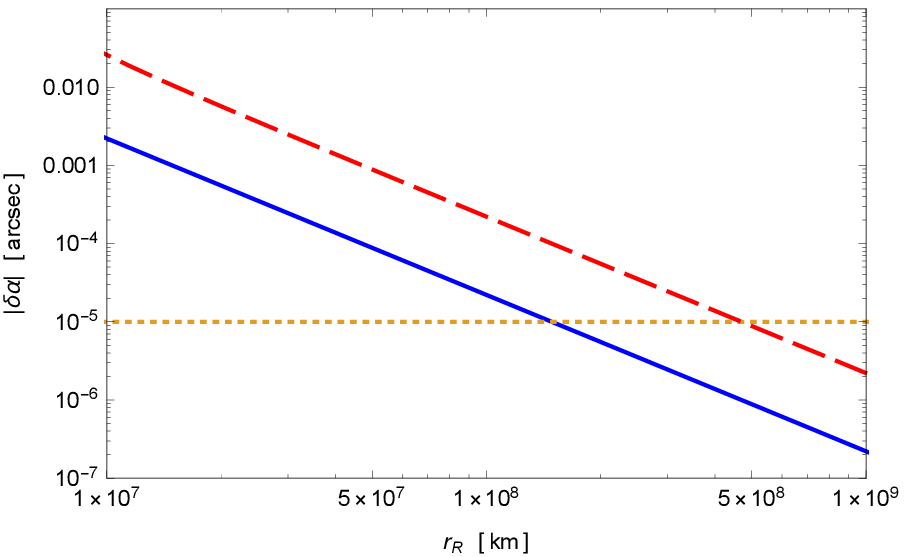}
\caption{ 
$\delta\alpha$ given by Eq. (\ref{alpha-Sun}) 
for the Sun. 
The vertical axis denotes the finite-distance correction 
to the deflection angle of light 
and the horizontal axis denotes the receiver distance $r_R$. 
The solid curve (blue in color) and dashed one (red in color) 
correspond to $b=R_{\odot}$ and $b=10 R_{\odot}$, respectively. 
The dotted line (yellow in color) corresponds to 10 micro arcseconds. 
}
\label{fig-Sun}
\end{figure}

Figure \ref{fig-Sun} shows numerical calculations 
of the bending angle of light 
due to the finite distance of the receiver. 
These analytic and numerical results, 
Eq. (\ref{alpha-Sun}) and Figure $\ref{fig-Sun}$, 
are consistent with each other 
and they suggest that $\delta\alpha$ by the solar mass 
might not be negligible in future astrometry observations. 
Note that 
the above correction happens to be comparable to 
the deflection of light at the second post-Newtonian order. 
If a 2PN test of the light bending by the Sun 
is done in the future, therefore, 
the above correction might be relevant.

\subsection{Sgr A$^{\ast}$}
Next, we consider the strong deflection limit. 
One of the most plausible candidates for the strong deflection 
is at the center of our Galaxy. 
It is identified with Sgr A$^{\ast}$. 
In this case, the receiver distance is much larger than 
the impact parameter of light, 
while a source star may be in the central region of our Galaxy. 

For Sgr A$^{\ast}$, Eq. (\ref{order-alpha}) implies 
\begin{align}
\delta\alpha 
&\sim 
\frac{Mb}{r_S{}^2} 
\nonumber\\
&\sim 
10^{-5} \mbox{arcsec.} 
\times 
\left(\frac{M}{4 \times 10^6 M_{\odot}}\right) 
\left(\frac{b}{3M}\right)
\left(\frac{0.1 \mbox{pc}}{r_S}\right)^2 , 
\label{alpha-Sgr}
\end{align}
where we assume the mass of the central black hole 
as $M \sim 4 \times 10^6 M_{\odot}$ 
and  
the strong deflection limit as $b \sim 3M$. 
This angle might be reachable in near-future astronomy.

\begin{figure}
\includegraphics[width=10cm]{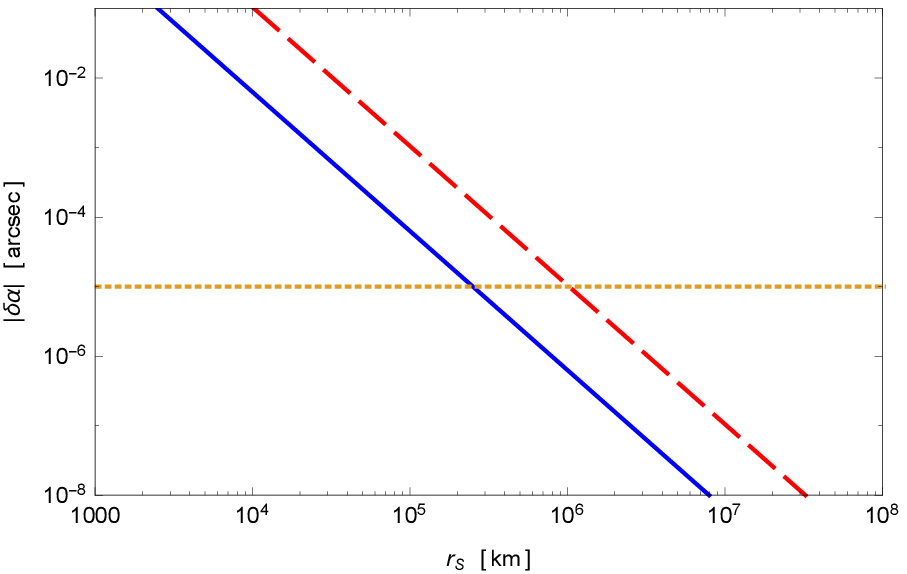}
\caption{ 
$\delta\alpha$ given by Eq. (\ref{alpha-Sgr}) 
for the Sgr A$^{\ast}$. 
The vertical axis denotes the finite-distance correction 
to the deflection angle of light 
and the horizontal axis denotes the source distance $r_S$. 
The solid curve (blue in color) and dashed one (red in color) 
correspond to $b=6 M$ and $b=10^2 M$, respectively. 
The dotted line (yellow in color) corresponds to 10 micro arcseconds. 
}
\label{fig-Sgr}
\end{figure}

Please see Figure \ref{fig-Sgr} for numerical calculations 
of the finite-distance correction due to the source location. 
This figure and also Eq. (\ref{alpha-Sgr}) suggest that 
$\delta\alpha$ 
can be of the order of ten (or more) micro arcseconds, 
if a source star is sufficiently close to Sgr A$^{\ast}$, 
for instance within a tenth of one parsec from Sgr A$^{\ast}$. 
In other words, 
for such a close source case, 
even though the source is still in the weak field, 
the infinite-distance limit is no longer sufficient.  
We have to take account of finite-distance corrections 
that are proposed in this paper.

\section{Conclusion}
For a static, spherically symmetric and asymptotically flat spacetime, 
we used the Gauss-Bonnet theorem to 
show that Eq. (\ref{alpha-full}) gives the exact bending angle of light 
especially by taking account of the finite distance from a lens object 
to a light source and a receiver,  
such that the definition can be valid also in the strong deflection limit. 
Finally, this method was applied to Schwarzschild spacetime 
as a concrete example. 
We discussed also possible observational implications. 
The proposed finite-distance corrections for Sgr A$^{\ast}$ for instance 
are able to amount to $\sim 10^{-5}$ arcseconds 
for a limited parameter range, 
which may be within the scope of near-future astronomy. 
Also the finite-distance correction to the Sun 
is $\sim 10^{-5}$ arcseconds, 
while this case is in the weak-field limit.

It would be interesting to apply Eq. (\ref{delta-alpha}) or 
Eq. (\ref{delta-alpha2}) to other black hole models 
and then to study a relation with future observations. 
It is left for future work.

\begin{acknowledgments}
We are grateful to Marcus Werner for the stimulating discussions, 
especially for his useful comments on their approach based 
on the Gauss-Bonnet theorem. 
We wish to thank Makoto Sakaki for giving us 
the useful literature information on the Gauss-Bonnet theorem. 
We would like to thank Toshifumi Futamase, Masumi Kasai, 
Yuuiti Sendouda, Ryuichi Takahashi, Atushi Naruko, Yuya Nakamura and 
Naoki Tsukamoto 
for the useful conversations. 
We would like to thank Hideki Ishihara for his hospitality 
at JGRG26 
(The 26th Workshop on General Relativity and Gravitation in Japan)  
in Osaka, 
where this work was largely developed. 
This work was supported 
in part by Japan Society for the Promotion of Science 
Grant-in-Aid for Scientific Research, 
No. 26400262 (H.A.) and 
in part by by Ministry of Education, Culture, Sports, Science, and Technology,  
No. 15H00772 (H.A.).
\end{acknowledgments}

\appendix
\section{Derivation of Eq. (\ref{alpha-approx})}
While the light ray passes near the photon sphere 
($r_0 \sim 3M$), 
we assume that the receiver and the source are very far from 
the lens ($r_R \gg r_0$ and $r_S \gg r_0$). 
Straightforward calculations for the Schwarzschild case give 
\begin{eqnarray}
\Psi_{R}-\Psi_{S}&=&\arcsin(bu_R)+\arcsin(bu_S)-\pi 
\nonumber\\
                & &-\frac{M bu_R^2}{\sqrt{1-b^{2}u_R^{2}}}
                   -\frac{M bu_S^2}{\sqrt{1-b^{2}u_S^{2}}}
                   +O\left(\frac{M^2}{r_R^2}, \frac{M^2}{r_S^2}\right) , 
\label{app-1}
\end{eqnarray}
\begin{eqnarray}
\int^0_{u_R}\frac{du}{\sqrt{F(u)}}+\int^0_{u_S}\frac{du}{\sqrt{F(u)}}
    &=&-\arcsin(bu_R)
       +\frac{2M}{b\sqrt{1-b^{2}u^{2}_R}}
       \left(1-\frac{1}{2}b^{2}u_{R}^{2}\right)\nonumber\\
    & &-\arcsin{(bu_{S})}
       +\frac{2M}{b\sqrt{1-b^{2}u_{S}^{2}}}
       \left(1-\frac{1}{2}b^{2}u_{S}^{2}\right)\nonumber\\
    & &-\frac{4M}{b}
       +O\left(\frac{M^2}{r_{R}^2}, \frac{M^2}{r_{S}^2}\right) , 
\label{app-2}
\end{eqnarray}
and 
\begin{eqnarray}
2\int^{u_0}_0\frac{du}{\sqrt{F(u)}}
    =2\log\left(\frac{12(2-\sqrt{3})r_0}{r_0-3M}\right)
     +O\left(1-\frac{3M}{r_0}\right) ,  
\label{app-3}
\end{eqnarray}
where we use Eqs. (\ref{sinPsi}) and (\ref{F-Sch}). 
Eq. (\ref{app-3}) was derived by Iyer and Petters \cite{Iyer}. 
By substituting Eqs. (\ref{app-1})-(\ref{app-3}) into Eq. (\ref{alpha-full}),  
we obtain Eq. (\ref{alpha-approx}). 
Note that $\arcsin(bu_R)$ and $\arcsin(bu_S)$, 
which are the leading terms of $\Psi_R$ and $\Psi_S$, 
cancel out with the terms in the path integrals 
of Eq. (\ref{app-2}). 
Therefore, $\arcsin(bu_R)$ and $\arcsin(bu_S)$ do not appear 
in Eq. (\ref{alpha-approx}), 
while $\Psi_R$ and $\Psi_S$ do contribute to $\alpha$.

\newpage

\newpage

\clearpage

\begin{thebibliography}{99}
\bibitem{comment-Bozza+}
In the present paper, we use the optical metric 
for suitably defining the angle in $M^{\mbox{opt}}$. 
On the other hand, Bozza and Scarpetta \cite{Bozza+} 
did not take care of this matter 
and they used the spatial part of the spacetime metric 
for its simplicity.
\end{thebibliography}
\end{document}